\documentstyle[aps,preprint]{revtex}
\begin{document}

\title{Temporal correlations versus noise in the correlation matrix formalism:
an example of the brain auditory response}

\author{J. Kwapie\'n$^{1}$, S. Dro\.zd\.z$^{1,2}$ and A.A. Ioannides$^{3}$}

\address{
  $^{1}$ Institute of Nuclear Physics, PL--31-342 Krak\'ow, Poland,\\
  $^{2}$ Institut f\"ur Kernphysik, Forschungszentrum J\"ulich,
  D--52425 J\"ulich, Germany,\\
  $^{3}$ Laboratory for Human Brain Dynamics, Brain Science Institute, RIKEN, Wako-shi,351-0198, Japan.}

\date{\today}

\maketitle

\begin{abstract}

  We adopt the concept of the correlation matrix to study correlations
  among sequences of time-extended events occuring repeatedly at consecutive 
  time-intervals. As an application we analyse the magnetoencephalography    
  recordings obtained from human auditory cortex in epoch
  mode during delivery of sound stimuli to the left or right ear. We look
  into statistical properties and the eigenvalue spectrum of the
  correlation matrix $\bf C$ calculated for signals corresponding to
  different trials and originating from the same or opposite hemispheres.
  The spectrum of $\bf C$ largely agrees with the universal properties of the
  Gaussian orthogonal ensemble of random matrices,  with deviations 
  characterised by eigenvectors with high eigenvalues.  The properties of 
  these eigenvectors and eigenvalues provide an elegant and powerful way of 
  quantifying the degree of the underlying collectivity during well defined 
  latency intervals with respect to stimulus onset.  We also extend this 
  analysis to study the time-lagged interhemispheric correlations, as a 
  computationally less demanding alternative to other methods such as mutual 
  information.

\end{abstract}

\newpage

\section{Introduction}

  Studying complex systems is typically based on analyzing large,
  multivariate data. Since, in general terms, complexity is primarily
  connected with coexistence of collectivity and chaos or even noise,
  it is of crucial importance to find an appropriate low dimensional 
  representation of an underlying high dimensional dynamical system.
  In many cases this aims at denoising and compressing dynamic imaging 
  data. Such a problem is particularly frequent in the area of the brain 
  research where a complex but relatively sparse connectivity prevails. 
  Understanding brain function requires a characterisation and quantification 
  of the correlations in the signals generated at different areas. 

  Direct pathways connect the sensory organs with the corresponding primary 
  cortical areas.  In the auditory system of interest here, delivery 
  of a stimulus to either the left or the right ear is relayed to both 
  primary auditory cortices, with stronger and earlier response on 
  the contralateral side.  The first cortical response arrives very early, 
  well within 20 milliseconds, but it is too weak to be mapped non-invasively 
  from outside.  Successive waves of cortical activation follow with the 
  strongest around 80-100 ms.  For a simple stimulus and no cognitive task 
  required the response as seen in the average is effectively over within 
  the first 200-300 milliseconds.  More elaborate analysis shows that the 
  "echoic memory" last for a few seconds~\cite{NR1,NR2}.  Furthermore 
  the activity in each area of the cortex, including the auditory cortex and 
  its subdivisions,  is determined by a plethora of interactions with 
  other areas and not just the direct pathway from the cochlea.  The 
  variability of the evoked response possibly reflects the many ways a given 
  input in the periphery can be modulated before the strong cortical 
  activations emerge~\cite{Liu98}.  Our treatment of the activity from each 
  auditory cortex as an independent signal bypasses this complexity by 
  lumping many effects into information theoretic measures.   The advantage 
  of this approach is that it leads to quantitative analysis of stochastic 
  and collective aspects of the complex phenomena in the auditory cortex and 
  the brain at large.

  In our previous work~\cite{Kwapien98} we have established the existence of 
  correlations between activity in the two auditory cortices, using mutual
  information~\cite{Fraser86} as a measure of statistical dependence.
  The analysis showed that collectivity and noise were present in the 
  data~\cite{Drozdz99a}.

  Usually, one analyzes a set of simultaneously recorded signals which emerge 
  from the activity of sub-components of the system.
  Consequently, the presence of correlations in
  such signals is to be interpreted as a certain sort of 
  cooperation among several or all of these sub-components.
  Though closely related, our present approach is somewhat different.
  Instead of studying many subsystems at the same time, we deal with two
  brain areas only and aim at identifying repetitive structures and their
  time-relations in consecutive independent trials of delivery of the stimulus.
  We thus construct the correlation matrix (which is a normalized version
  of the covariance matrix~\cite{Broomhead86,Liu96}) whose entries express
  correlations among all the trials that are delivered by experiment.
  The difference relative to a conventional use of the correlation matrix
  is that now the indices of this matrix are labeling different presentations 
  of the stimulus and not different subsystems. The resulting eigenspectrum
  is then expected to carry information about deterministic,
  non-random properties, separated out from the noisy background whose
  nature can also be quantified.

\section{Experiment and Data}

  The details of the experiment can be found in our earlier
  articles~\cite{Liu96,Liu98,Kwapien98}. Here, for completeness, we
  sketch briefly only the most important facts. Five healthy male 
  volunteers participated in the auditory experiment. We used
  2x37-channel, two-dewar MEG apparatus (each dewar covered the temporal
  area in one hemisphere) to measure magnetic field generated by the
  cortical electric activity~\cite{Hamalainen93}. The stimuli were 1 kHz
  tones lasting 50 ms each delivered in three runs to the left, right or
  both ears in 1 second intervals. The single trial of delivery of
  stimulus was repeated 120 times for each kind of stimulation. The
  cortical signals were sampled with 1042 Hz frequency.
  Pilot runs were used to place each dewar in turn so that both the positive 
  and negative magnetic field extrema were captured by the 37 channel array. 
  With such a coverage it is feasible to construct linear combinations of the 
  signals which act like virtual electrodes "sensing" the activity in the 
  auditory cortex~\cite{Liu98}.  This computation can be done at each
  timeslice of each single trial independently, thus building the
  timeseries for each auditory cortex for further analysis~\cite{Kwapien98}.

  Delivery of a sound stimulus or any change in the continuous stimulus
  causes a characteristic activity in the auditory cortex which is best 
  illustrated by averaging many such events~\cite{Creutzfeldt95}. 
  The (averaged) evoked potential, appears in both hemispheres and has a 
  form of several positive and negative deflections of the magnetic field. 
  The most prominent feature of the average is a high amplitude deflection 
  at about 80-100 ms after the onset of the stimulus (so called M100). The  
  details of the average evoked response are hardly visible in each single 
  trial, partly because of strong background activity, which is not related 
  to the stimulus  and partly because of the latency jitter introduced by the 
  many feed-forward and feed-back interactions that occur intermittently 
  between the periphery and the cortex. If as signal we consider what is 
  fairly time-locked to the stimulus onset then signal-to-noise ratio is much 
  improved by averaging the signal over all single trials.

  We will consider two runs,  corresponding to stimuli delivered to the left 
  and right ear.  Each run comprises $N=120$ 
  single trials, thus we have 120 signals for each hemisphere and each
  kind of stimulation. The signals are represented by the time series 
  $x^{L,R}_{\alpha} (t_i)$ of length of $T=1042$ time slices 
  $(i = 1,..., 1042, \alpha = 1,..., 120)$
  each evenly covering 1 second time interval. Since all the stimuli were
  provided in precisely specified equidistant instants of time, 
  all the series can be adjusted so 
  that the onset of each stimulus corresponds to the same time slice $i =
  230$. Each signal starts 220 ms before and ends 780 ms after the onset.
  A band pass filter was applied in the 1-100 Hz range.

  For a simple auditory stimulus and no cognitive task associated with it, the 
  average evoked response lasts for 200-300 ms;  this is also reflected in 
  our earlier  mutual information study of the signals~\cite{Kwapien98}.  
  Since other parts of each series are associated with activity which is not  
  time-locked to the stimulus,  the appearence of similar events in both 
  hemispheres and across trials results in correlations that are much stronger
  in the first few hundred millisecond.  The presence of correlations 
  and collectivity can not be excluded {\em a priori} from other periods and 
  it is therefore of considerable interest to compare two such intervals.  
  We have settle on two such intervals,  each with 250 timeslices:  the first 
  we call the Evoked Potential (EP) interval and it covers the first 250 
  timeslices after stimulus onset, i.e. 250 time slices $(i=231,480)$ 
  (2-241 ms);  this is the period where the average signal is strong.  
  The second interval we consider as baseline or background (B) and for
  this we choose the interval from 501 ms and ending 740 ms after the
  onset of the stimulus $(i=751,1000)$.  Since the time between stimuli
  is one second our choice avoids the time just before stimulus onset,
  when anticipation and expectation is high while being as far as
  possible from the stimulus onset.

\section{Correlation matrix analysis}

  For the two time-series 
  $x_{\alpha}(t_i)$ and $x_{\beta}(t_i)$ of the same length, $(i = 1,...,T)$
  one defines the correlation function by the relation
  
\begin{equation}
  C_{\alpha,\beta} = {\sum_{i} (x_{\alpha}(t_i)-\bar{x}_{\alpha})
                     (x_{\beta}(t_i)-\bar{x}_{\beta})
             \over {\sqrt{\sum_{i}{(x_{\alpha}(t_i) - \bar{x}_{\alpha})^2} 
             \sum_{j}{(x_{\beta}(t_j)- \bar{x}_{\beta})^2}}}},
\label{eq:cab}
\end{equation}
  where $\bar{x}$ denotes a time average over the period studied.
  For two sets of $N$ time-series $x_{\alpha}(t_i)$ each 
  $(\alpha, \beta=1,...,N)$
  all combinations of the elements $C_{\alpha,\beta}$ can be used 
  as entries of the $N \times N$ correlation matrix $\bf C$.
  By diagonalizing $\bf C$ 
  \begin{equation}
  {\bf C}{\bf v}^k = {\lambda}_k {\bf v}^k,
  \label{eq:diag}
  \end{equation}
  one obtains the eigenvalues $\lambda_k$ $(k=1,...,N)$ 
  and the corresponding eigenvectors ${\bf v}^k = \{v^k_{\alpha}\}$.

  In the limiting case of entirely random correlations the distribution
  $\rho_C(\lambda)$ is known analytically~\cite{Edelman98} and reads:

  \begin{equation}
  \rho_C(\lambda) = {Q \over{2 \pi \sigma^2}} 
  {\sqrt{(\lambda_{max} - \lambda)(\lambda - \lambda_{min})} \over \lambda}
  \label{eq:rho}
  \end{equation} 
  where 
  \begin{equation}
  \lambda_{min}^{max} = \sigma^2 (1 + 1/Q \pm 2 \sqrt{1/Q})
  \label{eq:lambda}
  \end{equation}
  with $\lambda_{min} \le \lambda \le \lambda_{max}$, $Q=T/N \ge 1$, and
  where $\sigma^2$ is equal to the variance of the time series 
  (unity in our case).

  For our present detailed numerical analysis we select two characteristic
  subjects (DB and FB) out of all five subjects who participated in
  the experiment. The background activity in both subjects does not
  reveal any dominant rhythm which, if present in two signals, may
  introduce additional, spontaneous correlations not related to the stimulus.  
  The signals of DB reveal a relatively strong EPs and a good signal-to-noise
  ratio. FB is somehow on the other side of the spectrum of subjects,
  as its EPs are small and hardly visible and the signals are dominated
  by a high-frequency noise which results in a poor SNR. 
  The signals forming pairs in eq.~(\ref{eq:cab}) may come either from the
  same or from the opposite hemispheres. The first possibility we term
  the {\it one-hemisphere} correlation matrix and the latter one the {\it
  cross-hemisphere} correlation matrix. The first matrix is, by
  definition, real symmetric and the second one must be real but, in
  general, it is not symmetric.
  
  An interesting global characteristics of the dynamics encoded in $\bf C$
  is provided by the distribution of its elements. An example for such
  a distribution is shown in Fig.~1 for the one-hemipshere correlation 
  matrix. As one can see in the background region (solid lines) the 
  distributions are Gaussian-like centered at zero. This implies that the
  corresponding signals are statistically independent to a large extent.
  A significantly different situation is associated with the evoked potential
  part of the signal. 
  The most obvious effect is that the centre of mass of the distribution is
  shifted towards the positive values. In this respect 
  there is also a difference between the subjects:
  the average value of elements for DB (approx. 0.35) is considerably
  higher than for FB (0.1). This indicates that the signals in FB are on
  average less correlated even in the EP region than the signals recorded
  from DB. This may originate from either a smaller amplitude of the
  collective response of FB's cortex or from a much smaller
  signal-to-noise ratio. For the cross-hemisphere correlation matrix
  the relevant characteristics are similar. The only difference is that
  the shifts (in both subjects) are slightly smaller.

  More specific properties of the
  correlation matrix can be analysed after diagonalazing $\bf C$.   
  The one-hemisphere correlation matrix 
  is real and symmetric and consequently all its eigenvalues are real. 
  The structure of their distribution is displayed in Fig.~2. 
  The eigenvalues are shown for several characteristic cases: 
  two subjects, the left and right hemispheres and two regions (EP and B).

  The structure of the eigenvalue spectra depends on the
  subject but first of all on the region of the signal. 
  There is a clear separation of the
  largest eigenvalue from the rest of the spectrum in the EP region in DB.
  This effect is much less pronounced for FB and considerably reduced in B. 
  This can be understood if we compare this result with Fig.~1.
  To a first approximation the distribution of elements in EP can be 
  described as a shifted Gaussian~\cite{Drozdz99b}: 

  \begin{equation}
  {\bf C} = {\bf G} + \gamma {\bf U},
  \label{eq:gu}
  \end{equation} 
  where ${\bf G}$ denotes a Gaussian matrix centered at zero and ${\bf U}$
  is a matrix whose entries are all unity. $\gamma$ is a real
  number $0 \le \gamma \le 1$. Of course, the rank of ${\bf U}$ is one and,
  therefore, the second term alone in eq.~(\ref{eq:gu}) develops only one
  nonzero eigenvalue of magnitude $\gamma$. Since the expansion coefficients
  of this particular state are all equal this assigns a maximum of collectivity
  to such a state. If $\gamma$ is significantly larger than zero the structure
  of $\bf C$ is predetermined by the second term in eq.~(\ref{eq:gu}).
  As a result the spectrum of $\bf C$ comprises one collective state with
  large eigenvalue. Since in this case $\bf G$ constitutes only a 'noise'
  correction to $\gamma {\bf U}$ all the other states are connected 
  with significantly smaller eigenvalues. In terms of the signals analysed 
  here the first component of (\ref{eq:gu}) corresponds to uncorrelated
  background activity and noise and the second one originates from the
  synchronous response of the cortex to external stimuli. 
  Similar characteristics of collectivity on the level of the correlation 
  matrix has recently been identified~\cite{Drozdz99b} in correlations 
  among companies on the stock market. 

  In relation to eq.~(\ref{eq:rho}) the presence of a strongly separated 
  eigenvalue is one obvious deviation which is consistent with the non-random
  character of the corresponding eigenstate. Further deviations can be 
  identified by comparing the boundaries of our calculated spectrum to
  $\lambda_{min}^{max}$ of eq.~(\ref{eq:lambda}). For $Q=T/N=250/120$ we
  obtain $\lambda_{min}=0.944$ and $\lambda_{max}=2.866$. Clearly, there
  are several eigenvalues more which are larger than $\lambda_{max}$. This
  may indicate that the corresponding eigenstates absorb a fraction of the 
  collectivity. However a closer inspection shows that also on the other
  side of the spectrum there are eigenvalues smaller than  $\lambda_{min}$
  and basically no empty strip between $0$ and $\lambda_{min}$ can be
  seen. By this our empirical distribution seems to indicate that an
  effective $Q$ which determines this distribution is significantly smaller 
  than $Q=T/N$. This, in turn, may signal that the information content in
  the time-series of length $T$ is equivalent to a significantly shorter
  time-series. This conclusion is supported 
  by the time-dependence of the autocorrelation function
  calculated~\cite{Drozdz99a} from our signals. It drops down relatively
  slowly and reaches zero only after 20-30 time-steps between consecutive
  recordings. Memory effects are present and hence neighboring recordings 
  are not independent;  this of course is not surprising because neural 
  activity in the brain has a finite duration (and 25-30 ms is an important 
  time scale) and there are plenty of time-delayed processes and interactions 
  which will produce activity in neighbouring times with shared information. 
  One could explicitly test whether this is a reason our calculated 
  $\rho_C(\lambda)$ deviates from the prediction of eq.~(\ref{eq:rho}) by 
  recomputing $\bf C$ with appropriately sparser time-series. Unfortunately, 
  the number of recordings covering the EP is too small for this. 
  Instead we perform the following analysis: we generate the new time-series
  $d_{\alpha} (t_i)$ such that 
  $d_{\alpha} (t_i) = x_{\alpha} (t_{i+1}) - x_{\alpha}(t_i)$, i.e., the
  time-series of differences. These destroy the memory effects and now the
  autocorrelation function drops down very fast.          
  Fig.~3 shows the density of eigenvalues of the correlation matrix 
  generated from $d_{\alpha} (t_i)$. Now the agreement with eq.~(\ref{eq:rho})
  improves and becomes relatively good already when every second 
  time-point $i$ from $d_{\alpha} (t_i)$ is taken, such that the total number 
  of them remains the same $(T=250)$.
  Taking more distant points, leaving out intermediate ones, drastically 
  reduces the correlation between the remaining successive points. 
  The above thus illustrates the subtleties 
  connected with the correlation matrix analysis of time-series. 
  Replacing our original time-series $x_{\alpha} (t_i)$ by $d_{\alpha} (t_i)$
  improves the agreement with eq.~(\ref{eq:rho}) but at the same time the
  collective state connected with EP dissolves. This is due to disappearance
  in $d_{\alpha} (t_i)$ of the memory effects present in 
  $x_{\alpha} (t_i)$. Therefore, in the following we return to our
  original time-series.

  Another statistical measure of spectral fluctuations is provided 
  by the nearest-neighbor spacing distribution $P(s)$. The corresponding
  spacings $s = \lambda_{i+1}-\lambda_i$ 
  are computed after renormalizing the eigenvalues in such a way 
  that the average distance between the neighbors equals unity. 
  A related procedure is known as unfolding~\cite{Brody81,Mehta91,Drozdz91}.
  Two characteristic and typical examples of such distributions 
  corresponding to EP and B regions are shown in Fig.~4 (for DB).
  While in both cases these distributions agree well with the 
  Wigner distribution which corresponds to the Gaussian orthogonal ensemble
  (GOE) of random matrices, some deviations
  on the level of larger distances between neighboring states are more visible
  in the EP than in the B region. This in fact is consistent with the presence 
  of larger eigenvalues in the EP case as shown in Fig.~2. 
  Interestingly, the bulk of $P(s)$ even here agrees well with GOE. 
  In order to further quantify the observed deviations
  we also fitted the histograms with the so-called Brody distribution         
  \begin{equation}
  P_r(s) = (1+r) a s^r \exp (-as^{(1+r)})
  \label{eq:brody}
  \end{equation}
  where
  $a = [\Gamma((2+r)/(1+r))]^{1+r}$.
  Depending on a value of the repulsion parameter $r$, this distribution
  describes the intermediate situations between the Poisson 
  (no repulsion, $r=0$) and the standard Wigner $(r=1)$ distribution (GOE).
  The best fit in terms of eq.~(\ref{eq:brody}) gives $r=0.95$ in the EP
  and $r=0.93$ in the B case, respectively. Thus we clearly see that the 
  measurements share the universal properties of GOE. A departure
  betraying some collectivity is nevertheless present in both B and EP
  intervals, but even in the EP interval the effect of the stimulus does
  not change this picture significantly: it results in one or at most few
  remote distinct states in the sea of low eigenvalues of the GOE type.
 
  In order to further explore this effect we look at the distribution 
  of the eigenvector components $v^k_{\alpha}$ for the same cases as in 
  Fig.~4. Fig.~5 displays such a distribution generated from eigenvectors
  associated to one hundred lowest eigenvalues (main panels of the
  Figure) calculated both for the EP (upper part) and B (lower part)
  regions. The result is a perfectly Gaussian distribution in both cases. 
  However, in EP a completely different distribution (upper inset)
  corresponds to the state with the largest eigenvalue. The charactersitic 
  peak located at around 0.1 documents that majority of the trials 
  contribute to this eigenvector with similar strength. This eigenvector
  is thus associated with a typical behavior of many single-trial signals. 
  The component values in the largest eigenvalue in B also deviate from a 
  Gaussian distribution (inset in the lower part of Fig.~5) although in 
  this case their distribution is largely symmetric with respect to zero. 
  This makes the two $k=120$ eigenvectors in B and EP regions approximately 
  orthogonal which indicates a different mechanism generating collectivity 
  in these two regions.        

  A more explicit way to visualise the differences among the eigenvectors 
  is to look at the superposed signals 
  \begin{equation}
  x_{\lambda_k} (t_i) = \sum_{\alpha=1}^{120} v^{k}_{\alpha}
  x_{\alpha} (t_i).  
  \label{eq:sup}
  \end{equation}
  For $k=120$, 119 and 75 these are shown in Fig.~6 using the eigenvectors
  calculated for the EP (middle panel) and for B (lower panel) regions.
  The signals corresponding to the largest eigenvalues $(k=120)$ develop
  the largest amplitudes in both cases. In the first case (EP) it very closely 
  resembles a simple average (upper panel) over all the trials.  In the second 
  case (B) long range correlations are clearly present, demonstrating that 
  there is more in the signal than the short latency correlations in EP. 
  The large eigenvalues in B also show a 
  degree of collectivity. When signals weighted by the eigenvectors with
  the highest eigenvalue in EP and B are compared we see that there is
  essentially no amplification in the other region (i.e. in the EP
  interval when the B-weighted signals are used). This provides another
  indication that different mechanisms are responsible for the
  collectivity at these two different latency ranges. Analogous effects of
  collectivity for $k=119$ are already much weaker and disappear
  completely as an example of $k=75$ shows.

  We now turn to the cross-hemisphere correlation function, obtained by 
  forming pairs in eq.~(\ref{eq:cab}) from the time-series representing 
  opposite hemispheres ($x_{\alpha}^L(t_i)$  with $x_{\beta}^R(t_i)$).  
  Introducing in addition a time-lag $\tau$ between such 
  signals~\cite{Kwapien98}, and dropping the rather obvious superscripts for 
the left and right hemisphere, we define a delayed correlation matrix

\begin{equation}
\label{cf_lagged}
  C_{\alpha,\beta}(\tau) = {\sum_{i} (x_{\alpha}(t_i)-\bar{x}_{\alpha})
                     (x_{\beta}(t_i+\tau))-\bar{x}_{\beta})
             \over {\sqrt{\sum_{i}{(x_{\alpha}(t_i) - \bar{x}_{\alpha})^2}
             \sum_{j}{(x_{\beta}(t_j+\tau))- \bar{x}_{\beta})^2}}}},
\ \ \ \ \ \alpha,\beta = 1,...,N.
\end{equation}

A similar cross-correlation time-lag function has been employed in the past 
to investigate across trials correlations, but because of the high 
computational load of an exhaustive comparison across different delays 
the analysis was restricted to the computation of the time-lagged 
cross-correlation between the average and individual single 
trials~\cite{Liu96}.
The spectral decomposition of the cross-correlation matrix provides a more 
elegant approach, requiring the solution of the $\tau$-dependent eigenvalue 
problem

\begin{equation}
  {\bf C}(\tau){\bf v}^k(\tau) = {\lambda}_k(\tau) {\bf v}^k(\tau),
  \ \ \ \ \ k = 1,...,N.
\label{cf_lag_diag}
\end{equation}
Since $\bf C$ can now be asymmetric its eigenvalues $\lambda_k$ can be complex 
(but forming pairs of complex conjugate values since $\bf C$ remains real) 
and in our case they generically are complex indeed. 
One anticipated exception may occur when similarity of the
signals in both hemispheres takes place for a certain value of $\tau$.
In this case $\bf C$ is dominated by its symmetric component and the effect,
if present, is thus expected to be visible predominantly on the largest 
eigenvalue. It is more likely to see this effect in the EP
region of the time-series. We thus calculate the cross-hemisphere
correlation matrix from the $T=250$-long subintervals of
$x_{\alpha}^L(t_i)$ and $x_{\beta}^R(t_i)$ covering the EPs.       
Fig.~7 presents the resulting real and imaginary parts of the largest 
eigenvalue as a function of $\tau$ for 
two subjects and two kinds of stimulation (left and right ear).
As it is clearly seen the large real parts are accompanied by
vanishing imaginary parts.
Based on this figure several other interesting observations are to be made.
First of all $\lambda_{max}(\tau)$ strongly depends on $\tau$ and reaches
its maximum for a significantly nonzero value of $\tau$. This reflects the
already known fact~\cite{Kwapien98} that the contralateral (opposite to the
side the stimulus is delivered) hemisphere responds first and thus the 
maximum of synchronization occurs when the signals from the opposite 
hemispheres are shifted in time relative to each other.  
(Here $\tau > 0$ means that the signal from the right hemisphere 
is retarded relative to the left hemisphere and the opposite applies 
to $\tau < 0$). Furthermore, the magnitude $(\tau \sim 10$ms) of the time-delay
estimated from locations of the maxima agrees with an independent estimate
based on the mutual information~\cite{Kwapien98}. 
Even a stronger degree of synchronization for DB relative to FB, as can be
concluded from a significantly larger value of $\lambda_{max}$ in the former
case, agrees with this previous study.     

Finally, Fig.~8 shows some examples of the eigenvalue distribution on the
complex plane. In the EP region the specific value of the time-delay 
$(\tau=7$ms, upper panel) corresponds to maximum synchronization between 
the two hemispheres for this particular subject. Here we see one strongly
repelled eigenvalue with a large real part $(\sim 36.5)$ and vanishing 
imaginary part. An interesting sort of collectivity can be inferred from
an example shown in the middle panel $(\tau=-40$ms) of Fig.~8. 
Here the largest eigenvalue is about a factor of 3 repelled more in the 
imaginary axis direction than in the real direction. 
This indicates that the antisymmetric part of $\bf C$ is dominating it 
which expresses certain effects of antisynchronization
(synchronization between the signals opposite in phase).   
In the B region, on the other hand, there are basically no such effects 
of synchronization between the two hemispheres and, consequently,
the complex eigenvalues are distributed more or less uniformly around (0,0)
as an example in the lowest panel of Fig.~8 shows. 

\section{Conclusions}
  
  The standard application of the correlation matrix formalism is to study
  correlations among (nearly) coincident events in different 
  parts of a given system. A typical principal aim of the related analysis 
  is to extract a low-dimensional, non-random component which carries some 
  system specific information from the whole multi-dimensional background
  activity. The advantage of the correlation matrix formalism is that
  it allows to directly relate the results to universal predictions of 
  the theory of random matrices.        
  The present study shows that the correlation matrix provides a 
  useful tool for studying the underlying mechanism which gives rise to 
  collectivity from a collection of events or signals sampled in different 
  regions. The brain auditory experiment considered here is one example 
  where there is a need for such an analysis. In this way we were thus 
  able to quantify the nature of the background brain activity in two 
  distinct periods which turns out to be 
  largely consistent with the Gaussian orthogonal ensemble of random matrices, 
  both in absence as well as in presence of the evoked potentials. 
  The analysis also allows to compare the degree of collectivity from the 
  properties of the eigenvectors with the highest eigenvalues. Crucially 
  the same analysis allows also a quantification of the degree of collectivity.        
  The beginnings of how the method can be extended to study correlations 
  between the two sources of signals was also outlined. In this case the correlation matrix is asymmetric and results
  in complex eigenvalues. An immediate application of such an extension 
  is to look at correlations among signals recorded in our experiment from
  the opposite hemispheres. Introducing in addition the time-lag between 
  the signals one can study the effects of delayed synchronization between
  the two hemispheres. The quantitative 
  characteristics of such synchronization remain in agreement with those
  found by other means~\cite{Kwapien98}.

\newpage
\begin{center}
{\bf FIGURE CAPTIONS}
\end{center}
{\bf Fig.~1.} Distributions of $C_{\alpha,\beta}$ for the one-hemisphere 
  correlation matrix. The upper panel corresponds to DB and the lower one
  to FB. The solid lines display such distributions evaluated in the
  regions beyond evoked activity (B) and the dashed lines in the EP
  region. \\
{\bf Fig.~2.} Structure of the eigenvalue spectra of the correlation matrices
  (one-hemisphere correlations) for the two discussed regions of the
  signals (evoked potential - EP, background activity - B) for DB (upper
  part) and FB (lower part). In each panel there are two spectra of
  eigenvalues, corresponding to the right hemisphere (circles) and the
  left one (triangles). The eigenvalues are ordered from the smallest to
  the largest. \\
{\bf Fig.~3} Density of eigenvalues of the correlation matrix calculated
  from the $T=250$ points of the time-series $d_{\alpha}(t_i)$ of
  increments of the original time-series $x_{\alpha}(t_i)$, i.e.,  
  $d_{\alpha}(t_i) = x_{\alpha}(t_{i+1}) - x_{\alpha}(t_i)$.
  In the lower panel every second point of $d_{\alpha}(t_i)$ is taken but the 
  number of such points is still 250.  
  The dashed line corresponds to the distribution prescribed by
  eq.~(\ref{eq:rho}). \\ 
{\bf Fig.~4.} Nearest-neighbor $(s)$ spacing distribution (histogram) of the 
  eigenvalues of $\bf C$ for subject DB. The upper panel corresponds to 
  the evoked potential (EP) region of the time-series and the lower panel
  to the background (B) activity part. The distributions have been created
  after unfolding the eigenvalues. The smooth solid curves illustrate the
  Wigner distribution and the dashed curves the best fit in terms of the
  Brody distribution. \\
{\bf Fig.~5.} Distribution of the eigenvector components ($v_{\alpha}^k$) 
  for EP (upper part) and B (lower part) regions (subject DB). The main
  panels correspond to one hundred lowest eigenvalues, while the insets
  show plots of the same quantity for the eigenvector corresponding to 
  $\lambda_{max}$ ($k=120$). For comparison, Gaussian best fits are
  also presented (dotted lines). (Note different scales in the Figure.) \\
{\bf Fig.~6.} The comparison of the signal obtained by simple average over
  all 120 trials (upper panel) and the signals obtained from 
  eq.~(\ref{eq:sup}) for both regions, EP (middle part) and B (lower
  part) for subject DB. Signals in the middle and lower panels denote
  superpositions for $k=120$ (solid line), $k=119$ (dashed line) and
  $k=75$ (dotted line). \\
{\bf Fig.~7.} $\lambda_{max}(\tau)$ calculated from the cross-hemisphere 
  correlation matrix. The upper part corresponds to DB and the lower part
  to FB. Both panels illustrate two kinds of stimulation: left ear (LE) 
  and right ear (RE). The solid lines denote the real part of
  $\lambda_{max}$ while the dashed and dotted ones its imaginary part.
  The sign of $\tau$ denotes retardation of a signal from the
  right hemisphere ($\tau > 0$) or the left one ($\tau < 0$). \\
{\bf Fig.~8.} Examples of the eigenvalue distribution of the
  cross-hemisphere correlation matrix for the right ear stimulation for 
  DB obtained from the EP region (upper and middle panels) and the B
  region (lower panel). All parts present the distributions on the
  complex plane. The eigenvalues for $\tau = 7$, which corresponds to the
  maximum of $\lambda_{max}(\tau)$ in Fig.~7, are shown in the upper panel 
  and the eigenvalues for $\tau = -40$ (corresponding to strong
  antisymmetry of {\bf C}) are presented in the middle one. A typical
  distribution of the eigenvalues in the B region is illustrated in the
  lower part. (Note different scale in the middle panel.)

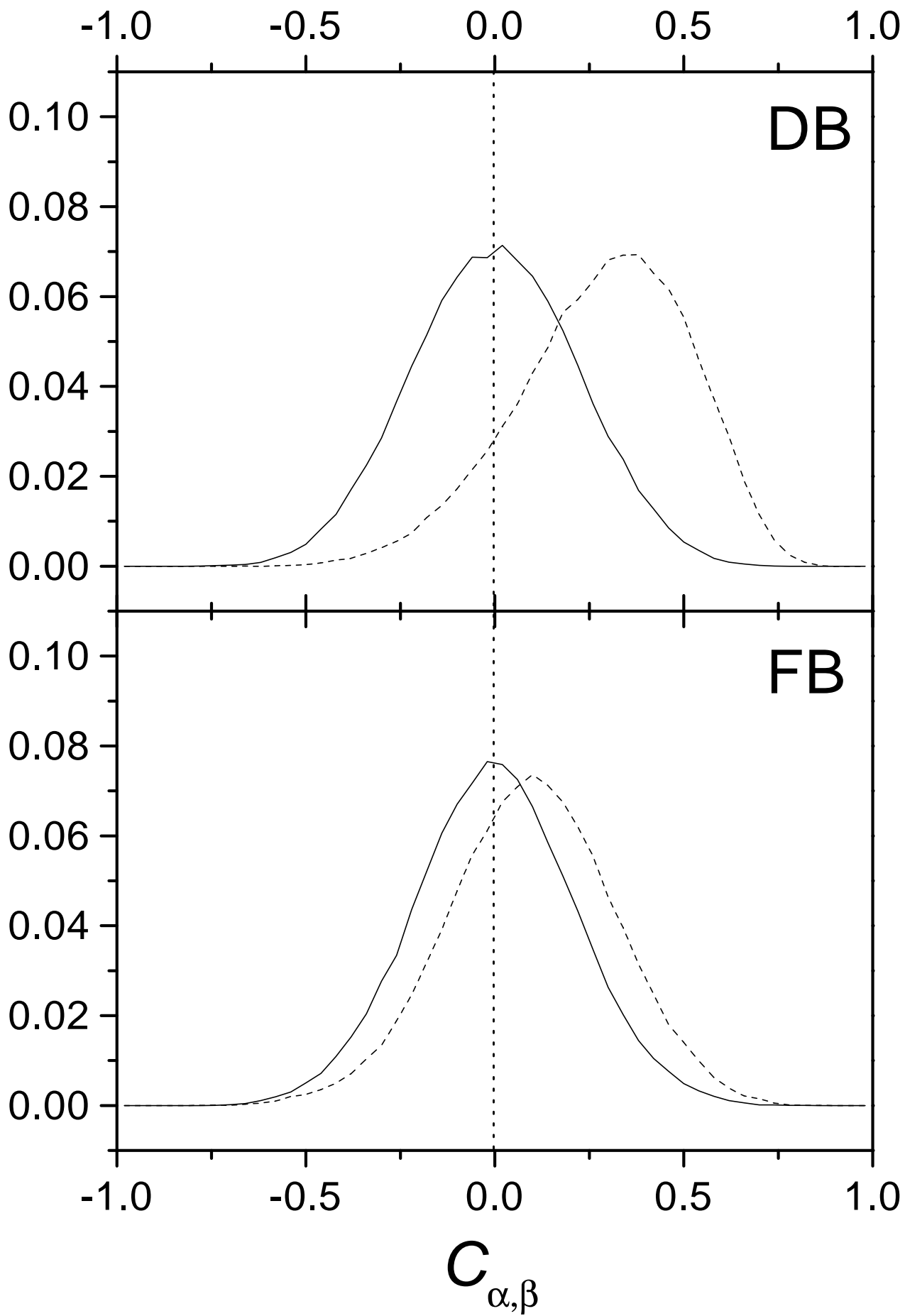
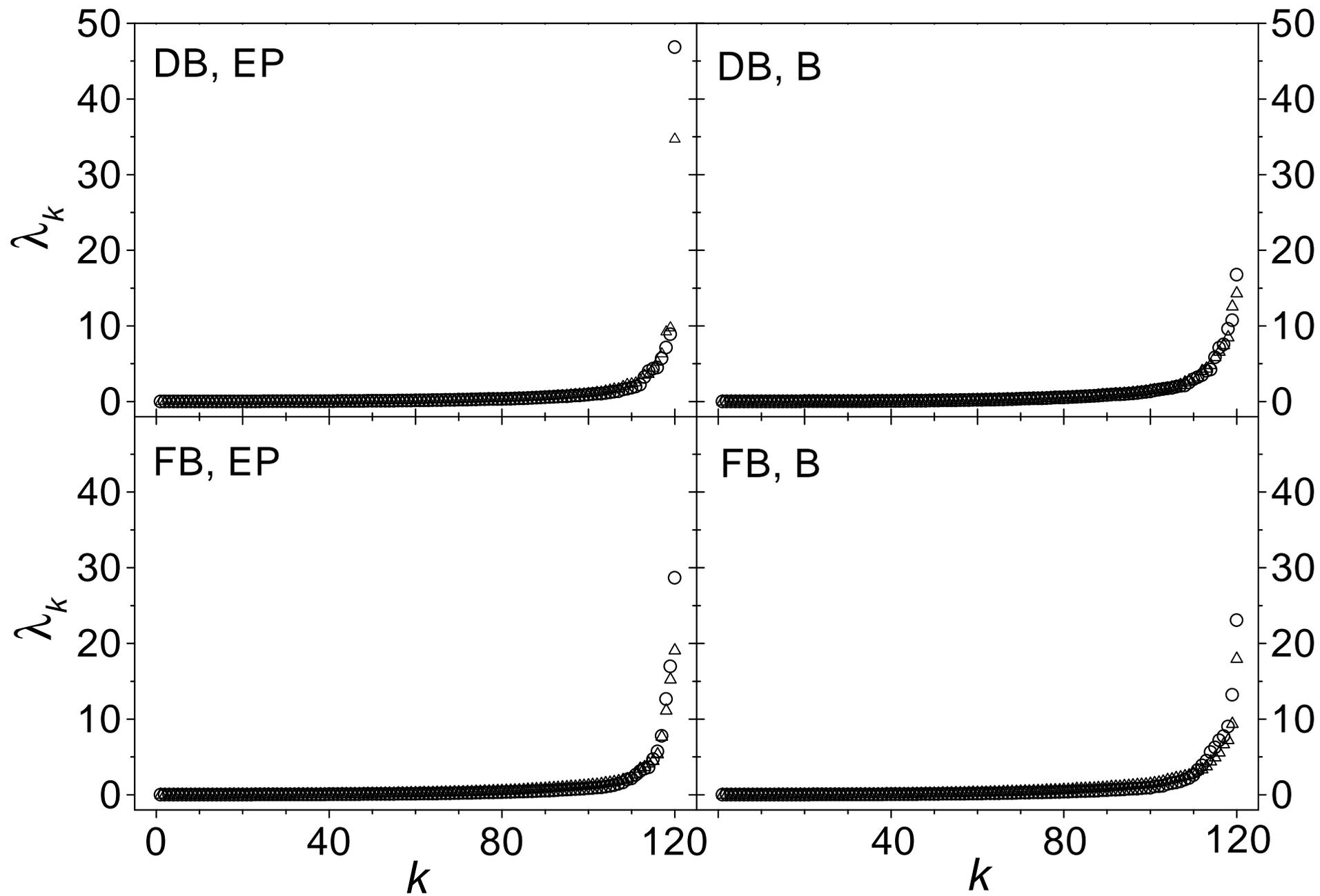
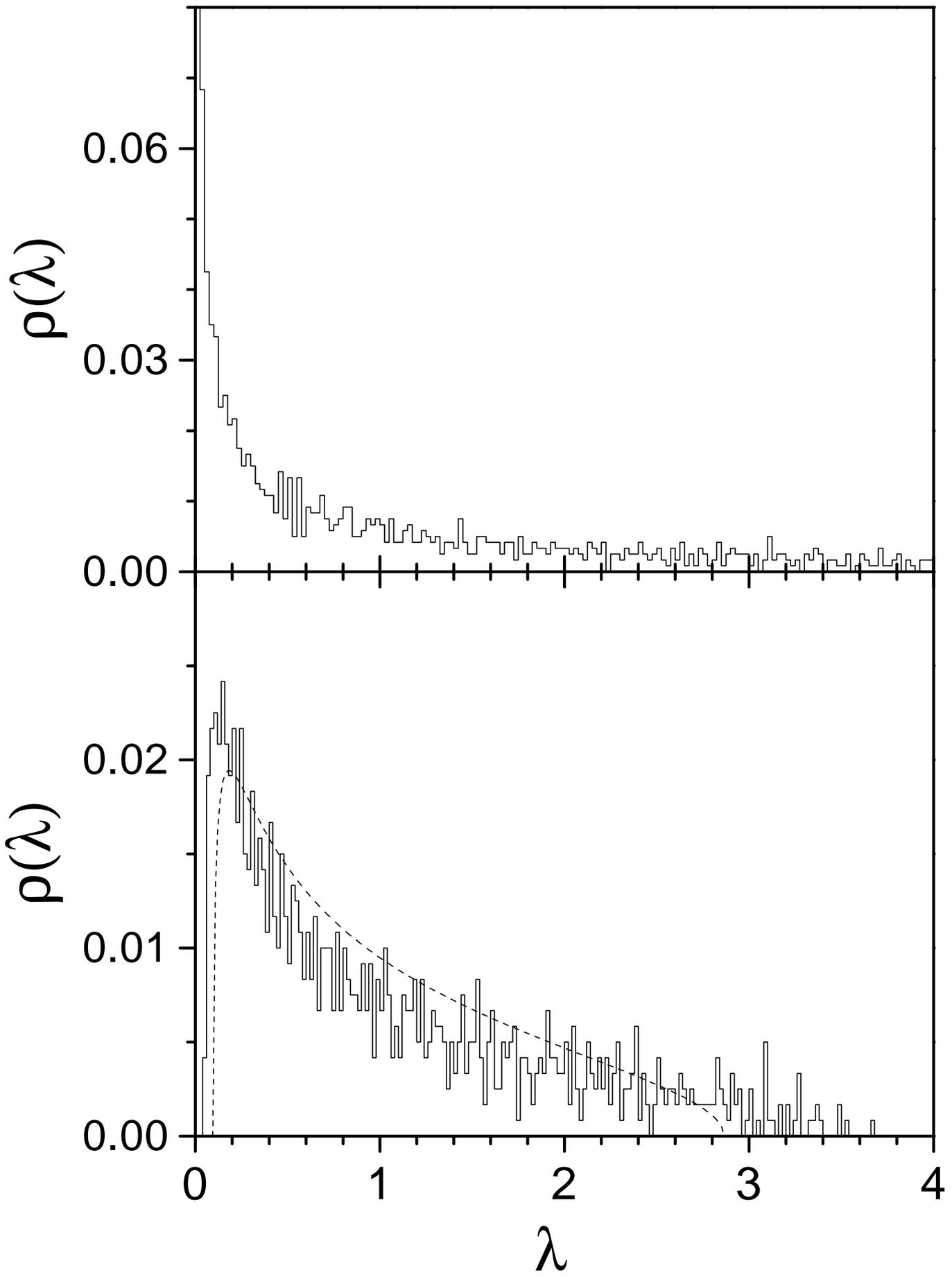
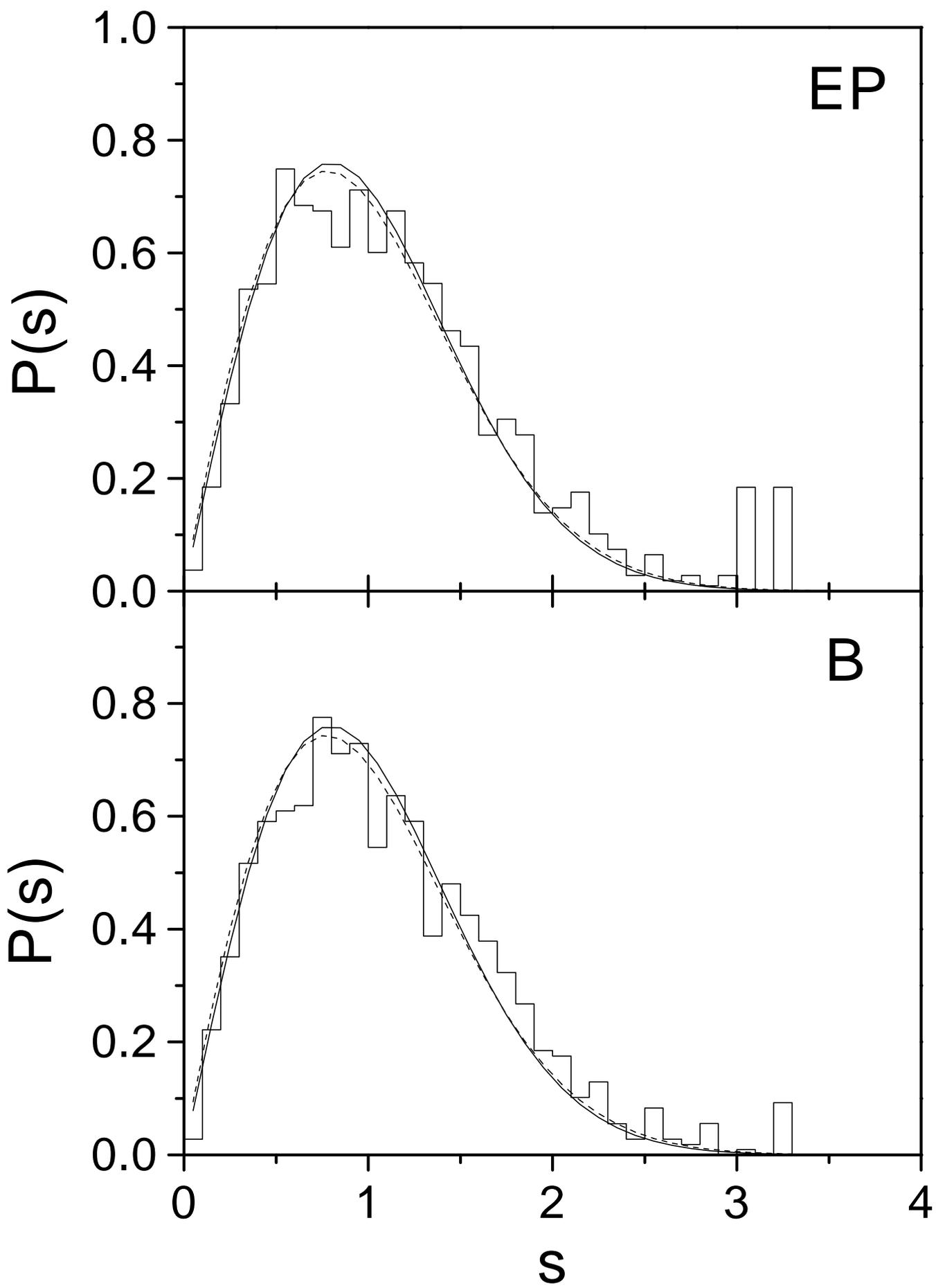
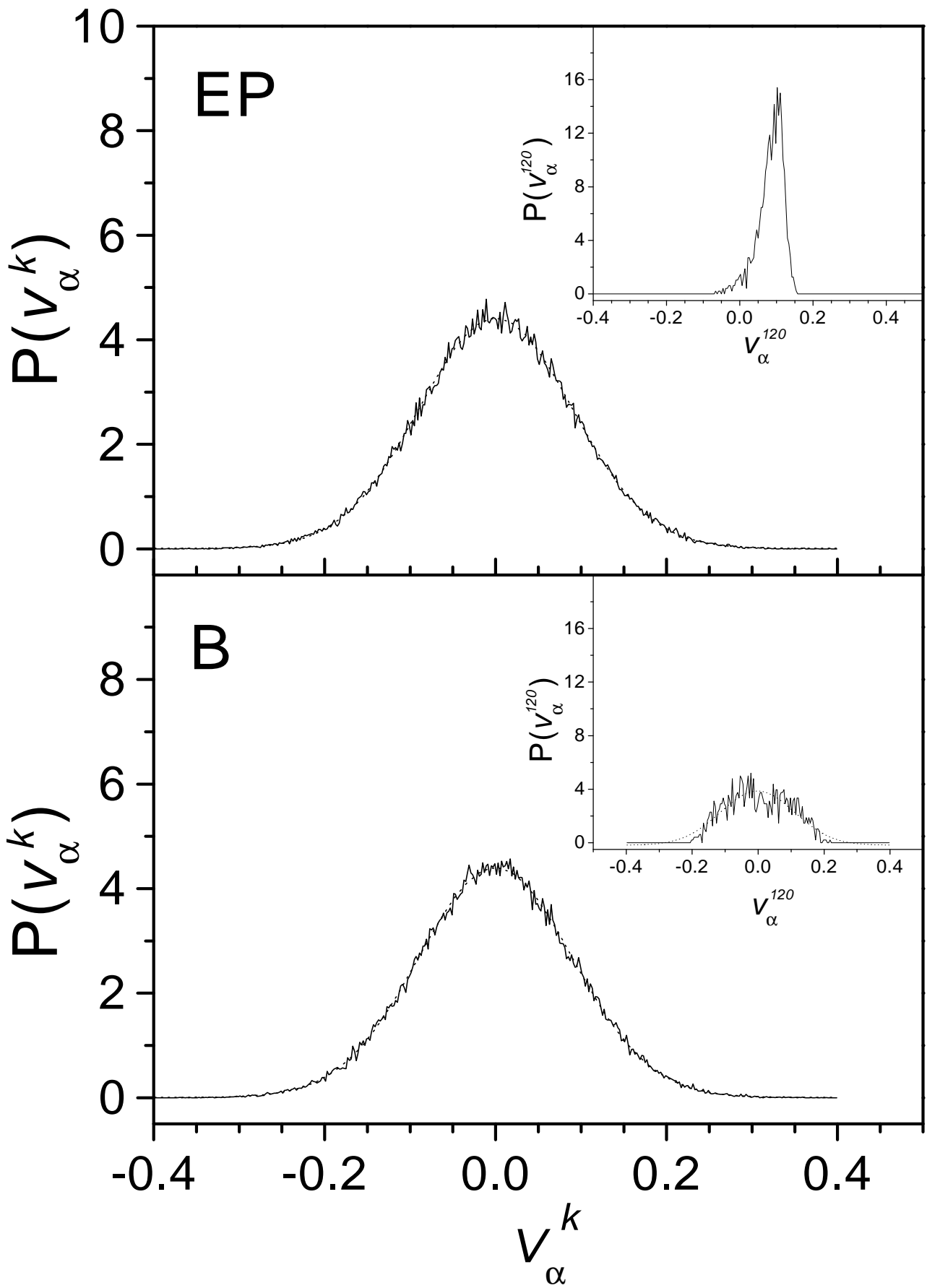
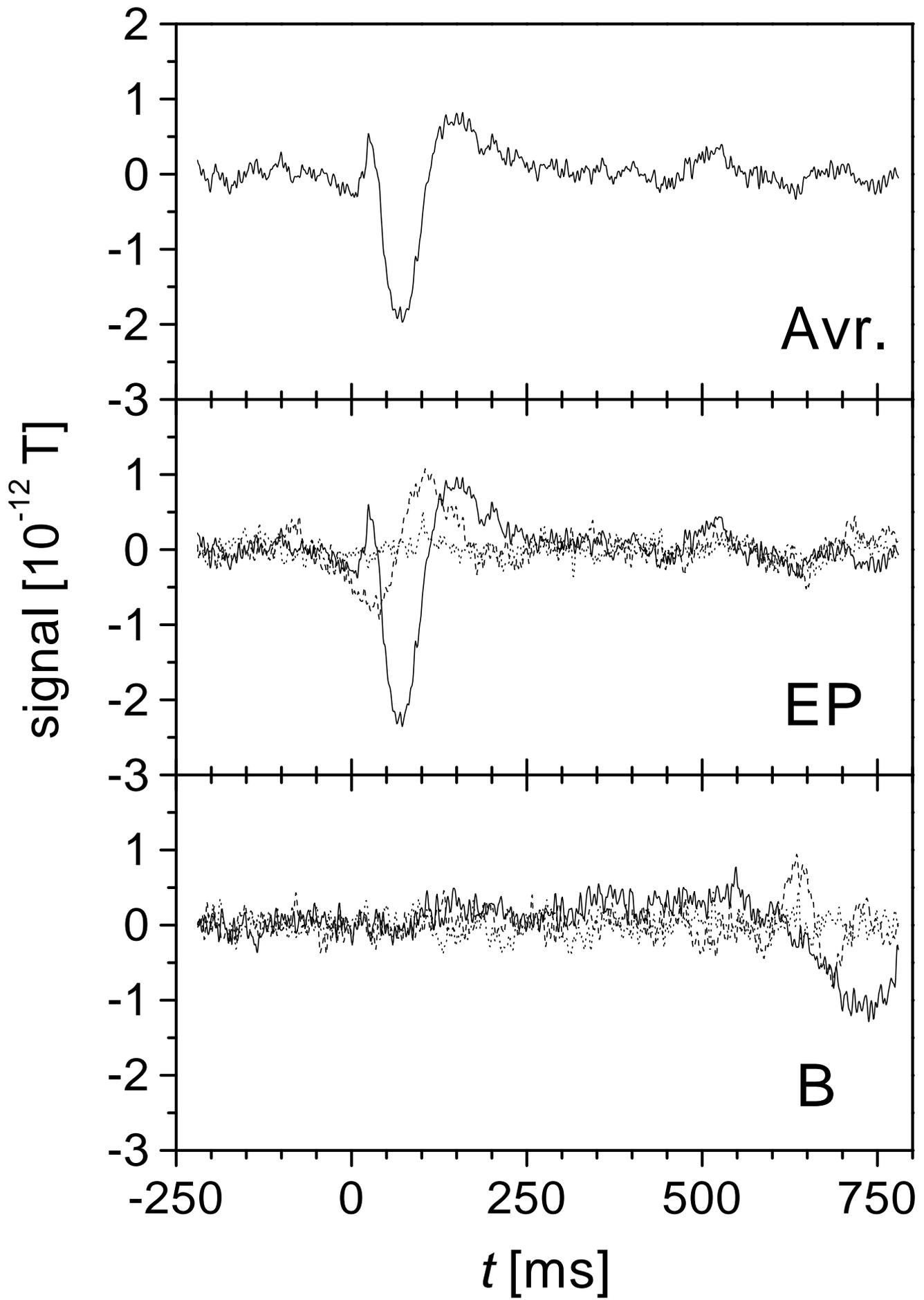
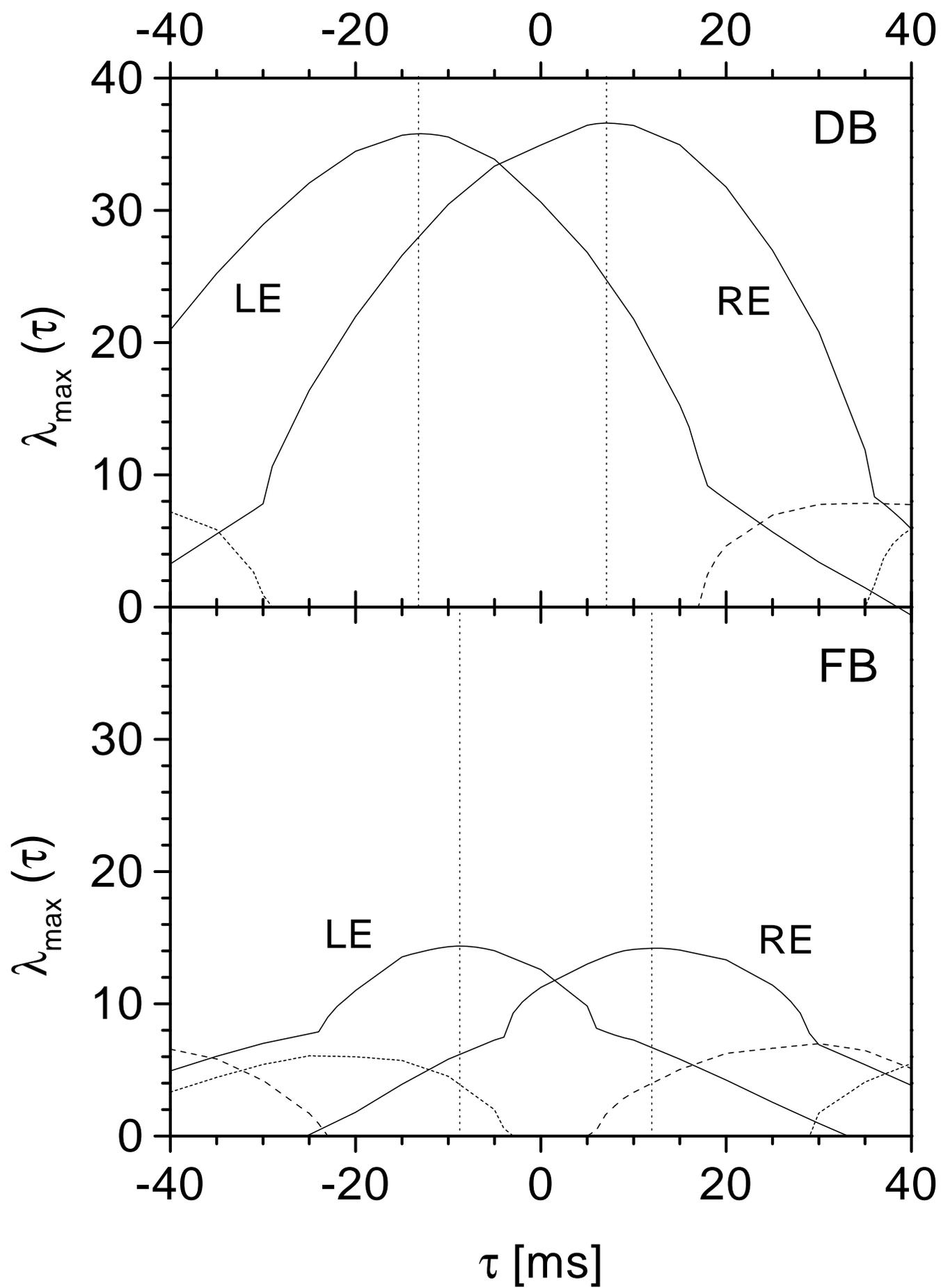
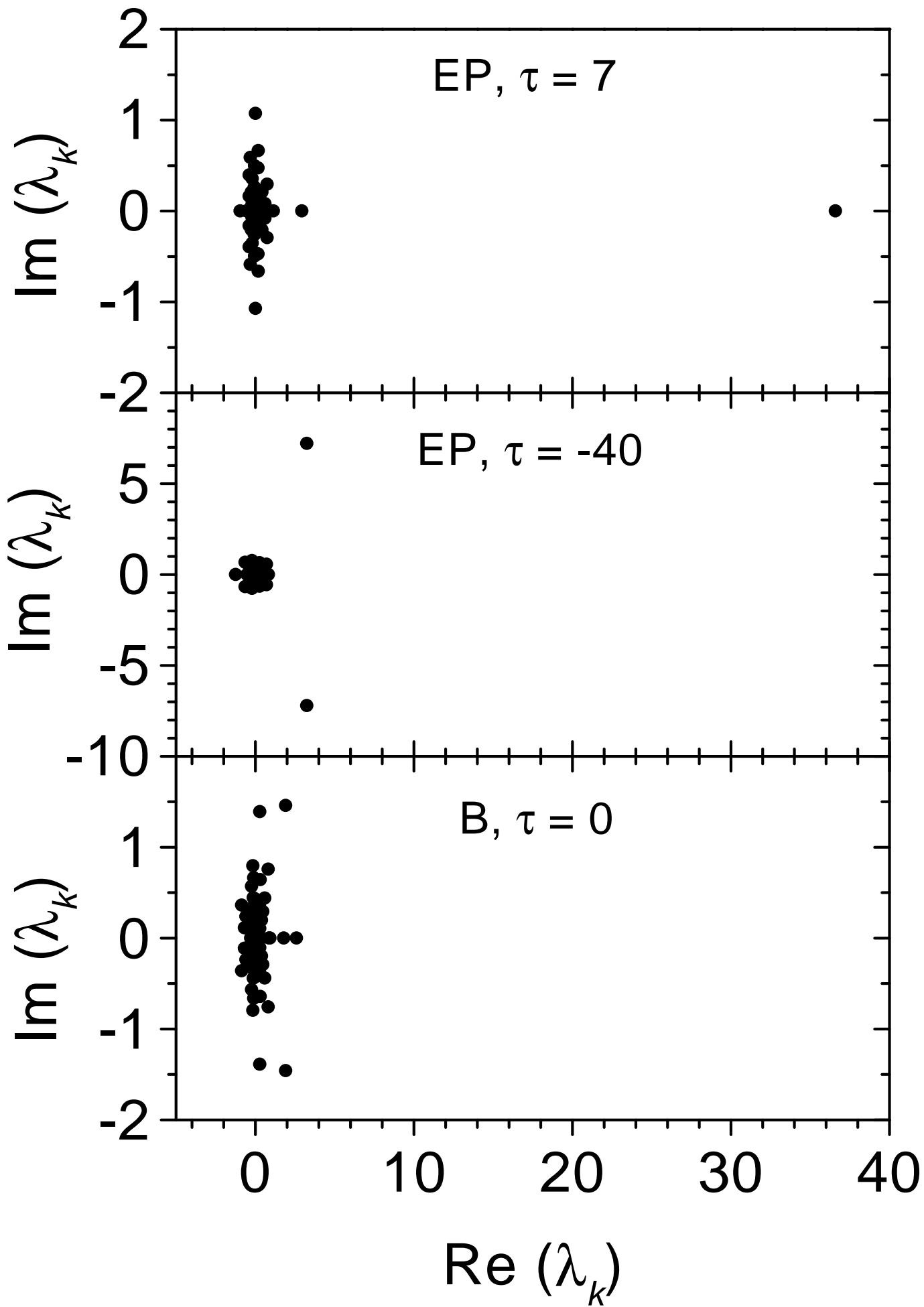
\end{document}